\documentclass{article}

\usepackage{arxiv}

\usepackage[utf8]{inputenc} 
\usepackage[T1]{fontenc}    
\usepackage{hyperref}       
\usepackage{url}            
\usepackage{booktabs}       
\usepackage{amsfonts}       
\usepackage{nicefrac}       
\usepackage{microtype}      
\usepackage{lipsum}		
\usepackage{graphicx}
\usepackage[sort,numbers]{natbib}
\usepackage{doi}

\usepackage{svg}
\graphicspath{{./img/}} 
\usepackage{amssymb} 
\usepackage{amsmath}
\usepackage{subfigure}
\usepackage{algorithm}
\usepackage{algpseudocode}

\title{The Theoretical Limits of Biometry}
\author{
	\href{https://orcid.org/0000-0002-9072-1535}{\includegraphics[scale=0.06]{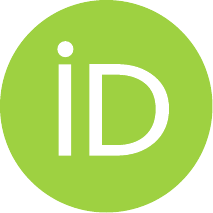}\hspace{1mm}Ga\"elle Candel} \\
	Ingenico Innovation Labs, Suresnes \\
	\texttt{firstname.lastname@ingenico.com} \\
}

\date{} 					

\renewcommand{\headeright}{}
\renewcommand{\undertitle}{}

\renewcommand{\shorttitle}{The Theoretical Limits of Biometry}

\hypersetup{
pdftitle={The Theoretical Limits of Biometry},
pdfsubject={cs.AI},
pdfauthor={Ga\"elle Candel},
pdfkeywords={Biometry, Identification, Template size, Modeling, Theory},
}

\begin{document}
\maketitle

\begin{abstract}
Biometry has proved its capability in terms of recognition accuracy.
Now, it is widely used for automated border control with the biometric passport, to unlock a smartphone or a computer with a fingerprint or a face recognition algorithm.
While identity \textit{verification} is widely democratized, pure \textit{identification} with no additional clues is still a work in progress.
The identification difficulty  depends on the population size, as the larger the group is, the larger the confusion risk.
For collision prevention, biometric traits must be sufficiently distinguishable to scale to considerable groups, and algorithms should be able to capture their differences accurately.

Most biometric works are purely experimental, and it is impossible to extrapolate the results to a smaller or a larger group.
In this work, we propose a theoretical analysis of the distinguishability problem, which governs the error rates of biometric systems.
We demonstrate simple relationships between the population size and the number of independent bits necessary to prevent collision in the presence of noise.
This work provides the lowest lower bound for memory requirements.
The results are very encouraging, as the biometry of the whole Earth population can fit in a regular disk, leaving some space for noise and redundancy.

\end{abstract}


\section{Introduction}

There are three ways to prove your identity or your ownership:

\begin{itemize}
\item \textit{Knowledge}: a password or a PIN,
\item \textit{Belongings}: a key (a physical object or a binary file), a smartphone, a (smart) card, a passport,
\item \textit{Being}: a physiological or behavioral trait: this is biometry.
\end{itemize}

Knowledge and belongings are strong evidence: they match or do not; there is no uncertainty.
One single character difference within a password prevents a user from logging in.
A password can be shared, but an object cannot be duplicated easily, unless it is digital.
The main problem with passwords is remembering several passwords with sufficient security.
Human brains need help remembering more than three passwords.
On the other hand, a password makes generating strong, diverse password easy but requires the file to be stored somewhere.
Computer failures make password inaccessible.

Biometry is an alternative to these two options: it is always with us and difficult to copy.
On the one hand, it prevents from sharing access to a sibling.
On the other hand, it prevents unauthorized users from claiming to be you, as making a counterfeit is very difficult.

The main problem of biometry is the accuracy of the systems.
First, selected traits must be permanent.
Hopefully, the iris, fingerprints, and veins are long-term attributes, unless surgery is done.
Next, we need accurate sensors and good algorithms.
During the last 20 years, there has been major progress in capturing devices where their capturing quality has improved.
These improvements have a direct impact by reducing noise, improving results' quality.
With a growing interest in biometrics, researchers have improved the accuracy of their algorithms, and were able to study their results over large populations.
These improvements have led to better performance and more reliable results.
However, there is still room for improvement, as error rates are not yet absolute zeros.

In Biometry, there are two ways to perform authentication:

\begin{itemize}
\item \textbf{Verification}, or $1:1$ authentication: the user claims "I am \texttt{X}", and the system compares the freshly captured probe to the stored reference of \texttt{X},
\item \textbf{Identification}, or $1:n$ authentication: the user only provides his biometric attribute. The system compares this probe to all stored references and returns the identity associated with the best matching reference.
\end{itemize}

Today, most research focuses on verification, a kind of "two-factor authentication": you need to claim your ID thanks to a smart card, an ID card, or by spelling it, \textit{and} to show one or several biometric attributes.
Therefore, you need to \textit{be} and \textit{have} the knowledge or a physical token to be authenticated.

In the ideal world, we no longer need to carry our ID cards or passports.
Our biometric attributes will be sufficient.
This is the identification target, recognizing you among a large list of candidates.
This task is much more difficult as there is no additional information
about the user. The traits and recognition algorithm should enforce sufficient distinguishability to prevent misrecognition.

The goal of this article is to study the limits of biometrics.
We provide simple relationships between the population size and the number of bits necessary to prevent collisions.
We additionally consider account the impact of noise, which greatly impacts recognition performances.

The article is structured in the following way:

\begin{itemize}

\item
  In the following part, we present some previous works that have been
  done to evaluate the distinguishability of a biometric system.
\item
  Next, we propose a theoretical analysis of the number of bits
  necessary to prevent collision between templates.
\item
  After that, we study the impact of noise on the recognition performances.
\item
  Then, we link our findings to traditional error rates by introducing a
  decision threshold.
\item
  We present the numerical results in a dedicated part.
\item
  Finally, we conclude this article and suggest research directions.
\end{itemize}

\hypertarget{related-works}{%
\section{Related Works}\label{related-works}}

Biometry is an experimental science, where most papers describe how data
are collected and processed and how to measure the distance between two
templates. One biometric system is compared to another one mainly by
looking at the error rates: the False Negative Rate and False Acceptance
Rate (FNR/FAR) for \emph{verification}, and the False Negative
Identification Rate and the False Positive Identification Rate (FNIR
/FPIR) for \emph{identification}. The negative and positive rate are
functions of a decision threshold, where if the acceptance threshold is
reduced, i.e., the difficulty is raised, then the negative rate
increases while the positive rate decreases.
It is common to share the equal error rate (EER), to compare two systems from
a fair perspective,
which is the value for which \(EER = FNR = FAR\), or equivalently
\(EER = FNIR = FPIR\) for identification.

These error rates are computed on a database with a specific size,
and the population size varies a lot from one study to another.
The problem is that the error rate \emph{depends} on the database size. If the
database is small, the average distance between templates is large.
However, the distance between templates is reduced when the database size increases, and the risk of making errors increases.
Therefore, the EER allows comparing two biometric systems only if the two databases are
approximately the same size.

To downplay the problem of generalization, researchers started to
study biometry from a more theoretical perspective. The first works have
been done on iris codes~{[}1{]},~{[}2{]}, recognized as one of
the most reliable biometrics. Iris codes are binary vectors of fixed
length, which are easy to study and manipulate. The authors measured the
Hamming distance between iris codes from different eyes and modeled the
distribution using binomial law. Using the observed mean and variance,
they estimated the number of degrees of freedom to \(244\) in the first
work and \(249\) in the second.

Degrees of freedom are synonymous with \emph{diversity}, i.e., the maximal
number of different possible patterns, which in the case of iris is
about \(2^{249} \approx 5.63 \times 10^{14}\) However, \emph{diversity}
does not mean ``maximal number of users.'' We need to take into account
the \emph{distinguishability}. Two users may have distinct iris codes.
However, this does not imply that iris codes are distinguishable. Indeed,
when performing several captures, some differences appear due to noise
(pupil dilatation, slight rotation of the head, analog to digital
conversion, light conditions, \ldots): for one user, we may observe
several close iris codes. Therefore, there must be a sufficient distance
between the two users to consider intra-variations. In these
former works, the authors studied the intra-distances between iris codes
from the same person and analyzed the impact of the decision threshold
on the FNR and FAR.

The works done on the iris were adapted to finger veins ~{[}3{]}. There are
different ways to process veins. One is to extract the location of vein
lines and crossing points as it is classically done for fingerprints.
Another way is to stay at the image level. In the work of ~{[}3{]},
pixels were classified into three categories: \texttt{vein},
\texttt{background} or \texttt{unknown}. Two images were compared pixel
by pixel, where a mismatch rate was computed. This rate was defined as
the ratio of non-overlapping vein pixels over the total vein pixels. Due
to the slight difference between the mismatch rate and Hamming distance, the
intra and inter distribution were modeled using a beta-binomial
distribution. They obtained results similar to iris codes, demonstrating
the suitability of finger vein patterns for a large population.

These previous works cannot be transposed to other biometrics easily, as
each biometric has its data structure. We need to define for each
case how distances between templates are measured and how distance
distributions are modeled. Computing the distances between templates is
often a trivial question, as the biometric system's matcher is
responsible for this task. However, the intra and inter-distribution modling must be done case by case.

Another example where the modeling differs from the two
previous examples is fingerprints. A fingerprint is often represented as
the list of minutiae with their location \((x, y)\) and angle
\(\theta\) ~{[}4{]}. In that case, the Hamming distance makes no sense.
Instead, the authors measured the probability of random correspondence
(PRC), which is the probability that two templates share a sufficiently
large number of minutiae. This work has an angle model, and
a model for minutiae location, and how the PRC is computed depends greatly
on this modeling. Results from one model to another vary greatly and
depend on the assumptions made. Fingerprints is one of the
oldest biometry, and many works have been done on this topic.
Nevertheless, there has yet to be a consensus on the PRC where, from one
work to another, the PRC ranges between \(10^{-80}\) and to \(10^{-11}\)
~{[}5{]}. These values are the two extremes, and most works suggest a
probability between \(10^{-60}\) and \(10^{-20}\). However, over time
and research progress, the value still changes from one study to another.

These previous works illustrate the heterogeneity of biometry and the
way error rates are computed.

A few works are trying to suggest a general methodology to compute
the \emph{biometric information} (BI), as if biometrics were a channel to
transmit information ~{[}6{]}. The BI is defined as ``\emph{the decrease
in uncertainty about the identity of a person due to a set of biometric
measurements}.'' Mathematically, it is defined as the Kullback-Leibler
divergence \(K(\mathbf{p}\| \mathbf{q})\) where \(\mathbf{p}\) is the
intra-distribution and \(\mathbf{q}\) the inter-distribution. The larger
the \(KL\) is, the more distinct those distributions and the easier it
is to recognize any individual. The KL divergence is also called the
\emph{relative entropy} as
\(K(\mathbf{p} \| \mathbf{q}) = H^*(\mathbf{p}, \mathbf{q}) - H(\mathbf{p})\),
where \(H\) is the Shannon entropy and \(H^*\) is the cross entropy. In
other words, the KL divergence represents the extra cost of encoding
\(\mathbf{p}\) using the codebook of \(\mathbf{q}\).

The BI aggregates the intra and
inter-distribution information into a single number, where there is no need to
define a threshold. This is another way to look at the problem of
distinguishability. This approach has two problems: the first
is the modeling of \(\mathbf{p}\) and \(\mathbf{q}\), which needs to be
solved case by case. The second, shared with other works like
~{[}1{]} and ~{[}3{]} is that getting an accurate intra-distribution
\(\mathbf{p}\) is difficult. Indeed, when we have a database with \(n\)
users and \(k\) images per user, we have
\(\frac{n \times (n-1)}{2} k^2\) pairs that can be used to compute
\(\mathbf{q}\). However, for \(\mathbf{p}\), we only have
\(n \times \binom{k}{2}\) pairs of images. In other words, we have
around \(n\) times less pairs to compute \(\mathbf{p}\) than for
\(\mathbf{q}\).

In the former work of ~{[}6{]}, BI is computed using features. Modeling
templates in a high dimensional space is difficult, and sometimes, their
dimensionality exceeds the number of samples, making the direct use of
PCA impossible. To remove the problem of high dimensionality, ~{[}7{]}
and ~{[}8{]} suggest looking at score or distance distributions to
facilitate the computation. They also suggest using a nearest neighbors'
estimator to improve the computation of \(\mathbf{p}\) using a few
samples. These notions of biometric information can easily be
generalized to multimodal biometrics ~{[}9{]}. The \emph{normalized
relative entropy} was derived from the BI by ~{[}10{]}, where the normalization aims
to estimate how much the considered system can be improved.

BI, degrees of freedom, and PRC are not directly equivalent. However,
they represent the limit of the biometric system considered. When
looking at numerical values, there is a large gap between the results
obtained from one approach to another. While ~{[}2{]} claims \(249\)
degrees of freedom for iris, ~{[}11{]} \(723\) for finger veins,
~{[}6{]} \(46.9\) bits for face recognition, other works claim much
lower values, such as \(12.62\) bits for face recognition and \(12.67\)
for fingerprints ~{[}9{]}, and only 10 for iris ~{[}7{]}.
Due to the authors' lack of diversity, it is unclear if these
differences are due to the approach selected or the accuracy of the
processing and/or the database quality.

Rather than evaluating another biometric system where our results would
be correlated to our ability to create a strong and accurate biometric
system, this work is purely theoretical. We propose analyzing of the
minimal number of bits necessary to prevent collision and confusion as a
function of the population size. Because of its simplicity, we focus on
templates as arrays of random uncorrelated bits, as for iris codes. We
consider noise in the matching process, as two captures are
always different.
The binary array model does not fit all biometric systems.
The uncorrelated bits hypothesis might be far
from the reality where correlations between features exist and people
from the same family may have similar traits.
Nonetheless, this model provides a lower bound for template size due to these ideal conditions.
This gives us an estimate of the achievable maximal compression level and minimal storage requirements.

\hypertarget{the-birthday-paradox}{%
\section{\texorpdfstring{The Birthday Paradox
\label{sec:nonoise}}{The Birthday Paradox }}\label{the-birthday-paradox}}

\hypertarget{a-short-reminder}{%
\subsection{A Short Reminder}\label{a-short-reminder}}

\begin{quote}
\emph{What is the probability that at least two persons in the room were
born the same day?}
\end{quote}

When asking this question, most people would say this probability
is very low. However, in a room of 30 people, this probability is
already close to \(70\%\).

In biometry, we have \(N\) users enrolled in a database, each
represented by a template, which is information extracted (most of the
time) from a raw image. These templates are supposed to be random;
sometimes, two users would get the same templates by chance. In that
case it will be impossible to distinguish the two users. We want to find
how to compute the template collision probability to prevent this event
from happening, which is the same thing as the birthday paradox.

To compute the collision probability, the simplest way is to compute the
probability of the complementary event, i.e., the probability that no one is
born on the same day. Assuming there are \(N\) persons and \(D\) days, we
have:

\begin{itemize}
\item
  User \(1\) has \(D\) choices for his birthday,
\item
  User \(2\) has \(D-1\) choices for his birthday,
\item
  \ldots{}
\item
  User \(N\) has \(D-N+1\) choices.
\end{itemize}

Therefore, there are several different ways to select birthdays are:

\[D \times (D-1)\times ... \times (D-N+1)= \frac{D!}{(D-N)!}\]

Next, we need to remember that each selected date is equiprobable,
with a probability of \(\frac{1}{D}\). Therefore, the probability that two
persons in a room were born on the same day is:

\begin{equation}\label{eq:birthday}
\begin{array}{ll}
  P(\text{At least 2 born the same day}) & =1 - P(\text{No one born the same day}) \\
    & = 1 - \frac{D!}{(D-N)!} \times \frac{1}{D^N}
    \end{array}
\end{equation}

\hypertarget{computing-the-collision-probability-with-large-numbers.}{%
\subsection{Computing the Collision Probability with Large
Numbers.}\label{computing-the-collision-probability-with-large-numbers.}}

We cannot play on the number of days in a year, but we can play on the
template length, denoted \(k\). Therefore, we denote \(T = 2^k\)
the total number of possible patterns which replaces \(D\) in the
previous equation Eq. \ref{eq:birthday}.

The population on Earth is currently close to \(10^{10}\).
The previous equation Eq. \ref{eq:birthday} contains factorials.
Computing factorials up to \(1000!\) is fine. However, computing factorials up to
\(10^{10}\) (and even above) accurately is intractable without
approximation. Therefore, in this section, we provide an upper and lower
bound to Eq. \ref{eq:birthday} to capture the main behavior.

With a product of large numbers, the most intuitive thing to do is to
move to the log scale: We obtain
\(\log\left(\frac{T!}{(T-N)!}\right) = \sum_{i=T-N+1}^T \log (i)\) which
cannot be more simplified. We can bound it by the left and by the right
with integrals:

\begin{equation}\label{eq:bounds}
    \int_{T-N}^T \log(x) dx \leq \sum_{i=T-N+1}^T \log(i) \leq \int_{T-N+1}^{T+1} \log(x)dx
\end{equation}

Analytical solution of the \(\log\) integral is obtained by integrating
by part:

\begin{equation}\label{eq:integral}
    \int_a^b \log(x) dx = \left[x \log x\right]_a^b - \int_a^b \frac{x}{x} dx = \left[x (\log x - 1)\right]_a^b
\end{equation}

\hypertarget{lower-bound-f_lowt-n}{%
\subsubsection{\texorpdfstring{Lower Bound
\(f_{low}(T, N)\)}{Lower Bound f\_\{low\}(T, N)}}\label{lower-bound-f_lowt-n}}

To get the lowest bound of Eq. \ref{eq:bounds}, we set \(a = T - N\) and
\(b = T\) in Eq. \ref{eq:integral}, leading to:

\[
\begin{array}{ll}
  f_{low}(T, N) & = T (\log T - 1) - (T-N) \left(\log(T-N) - 1\right) \\
    & = T \log T - (T - N)\log(T - N) - N
    \end{array}
\]

We can subtract the \(\frac{1}{T^N}\) part to get the \(\log\) of the
probability:

\[
\begin{array}{ll}
  \log\left(P_{low}(\text{No collision})\right) & = f_{low}(T, N) - N \log(T) \\
    & = (T - N) \log \frac{T}{T - N} - N \\
\end{array}
\]

We can get the two partial derivatives:

\[
\begin{array}{ll}
  \frac{\partial P_{low}}{\partial N} &= -\log\frac{T}{T-N} + \frac{T-N}{T-N} - 1 \\
  &= -\log \frac{T}{T-N} \\
  &= \log\frac{T-N}{T} < 0
\end{array}
\]

\[
\begin{array}{ll}
  \frac{\partial P_{low}}{\partial T} &= \log\frac{T}{T-N} + (T-N)\left[\frac{1}{T} - \frac{1}{T-N}\right]\\
      &= \log \frac{T}{T-N} + \frac{T-N}{T} - 1 \\
      &= \log \frac{T}{T-N} + 1 - \frac{N}{T} - 1\\
      &= \log \frac{(T - N) + N}{T-N} - \frac{N}{T} \\
      & \approx \frac{N}{T-N} - \frac{N}{T} = \frac{N^2}{T(T-N)} > 0
\end{array}
\]

Assuming that \(T\) is fixed, if \(N\) increases, then the collision
probability increases (the no-collision probability decreases), and
inversely, when \(N\) is fixed and \(T\) increases, the collision
probability decreases, which follows the natural intuition.

\hypertarget{upper-bound-f_upt-n}{%
\subsubsection{\texorpdfstring{Upper Bound
\(f_{up}(T, N)\)}{Upper Bound f\_\{up\}(T, N)}}\label{upper-bound-f_upt-n}}

We can re-do the same exercise with the upper bound, to ensure the
behavior is the same on both sides. Here, we update Eq. \ref{eq:integral}
by setting \(a = T - N+1\) and \(b = T + 1\):

\[
\begin{array}{ll}
  f_{up}(T, N) & = (T+1) (\log (T+1) - 1) - (T-N+1) \left(\log(T-N+1) - 1\right) \\
    & = (T+1) \log (T+1) - (T - N+1)\log(T-N+1) - N
    \end{array}
\]

Last, we subtract the \(\frac{1}{T^N}\) part:

\[
\begin{array}{ll}
 \log\left(P_{up}(\text{No collision})\right) & = f_{up}(T, N) - N \log(T) \\
   & = (T+1) \log (T+1) - (T - N+1)\log(T-N+1) - N  - N \log T \\
   & = (T+1) \log \frac{T+1}{T - N + 1} - N \log \frac{T}{T-N+1} - N
   \end{array}
\]

Deriving the two equations leads to very similar results.

\hypertarget{error-analysis}{%
\subsubsection{Error Analysis}\label{error-analysis}}

To see how well this approximation does, we can measure the difference
between \(f_{up}\) and \(f_{low}\), as the \(\log(\frac{1}{T^N})\) would
not impact the result:

\[
\begin{array}{ll}
  \Delta &= f_{up} - f_{down} \\
          &= \log(T+1) + T \log \frac{T+1}{T} - \log(T-N+1) - (T-N) \log\frac{T-N+1}{T-N} \\
          &= \log\frac{T+1}{T-N+1} + T \log \frac{T+1}{T} - (T-N) \log\frac{T-N+1}{T-N}
\end{array}
\]

To simplify this expression, we will approximate the \(\log\),
re-write them as \(\log(1 + x)\) where \(x\) is small relatively close
to \(0\). In that case, the \(\log\) terms can be approximated using:

\[\lim_{x \rightarrow 0} \log(1 + x) = x - \frac{x^2}{2} + \frac{x^3}{3} - ....\]

The first order is sufficient in our case. Therefore the terms are
approximated as follows: \[
\begin{array}{lll}
  \log \frac{T+1}{T-N+1} &= \log\left(1 + \frac{N}{T-N+1}\right) &\approx \frac{N}{T-N+1} \\
  \log \frac{T+1}{T} &= \log\left(1 + \frac{1}{T}\right) &\approx \frac{1}{T} \\
  \log \frac{T-N+1}{T-N} &= \log\left(1 + \frac{1}{T-N}\right) &\approx \frac{1}{T-N}
 \end{array}
\] For the term \(\frac{N}{T-N+1}\), the approximation is valid only if
\(T \gg N\), which can be verified experimentally (for \(T=365\) and
\(N=30\), \(T\) is already ten times larger than \(N\)).

Replacing the logs in the previous equation, we obtain:

\[
\begin{array}{ll}
  \Delta &= \frac{N}{T-N+1} + T \frac{1}{T} - (T-N) \frac{1}{T-N} \\
  &= \frac{N}{T-N+1} + 1 - 1 \\
  &= \frac{N}{T-N+1}
   \end{array}
\]

So, the greater \(T\) is relative to \(N\), the lower the difference
between the upper and the lower bound. As a reminder, \(\Delta\) is the
difference between the log probabilities.
Therefore, \(\exp(\Delta)\) is the ratio between the upper and lower bound,
which we expect to be close to \(1\). If \(T = xN -1\) where \(x\) is a positive value,
then \(\Delta = \frac{1}{x}\). If we suppose that \(x=100\), then
\(\exp(\Delta) = 1.01\). In other words, there is a difference of
\(1\%\) between the upper and the lower bound, which is an acceptable
error level.

Because of this small approximation rate, we will work on the lower bound in the rest of
this section, as its expression is much simpler to
manipulate.

\hypertarget{preventing-collision}{%
\subsection{Preventing Collision}\label{preventing-collision}}

Intuitively, the number of patterns \(T\) needs to be adapted to the
number of users \(N\). To simplify the problem and the analysis, we
express \(T\) as a function of \(N\) by setting \(T = xN\) where
\(x \in \mathbb{R}^{+*}\). After replacing \(T\) by \(xN\), the lower
bound becomes:

\[
\begin{array}{ll}
  \log\left(P_{low}(\text{No collision})\right)(N, x) & = (xN - N)\log\frac{xN}{xN - N} - N \\
                                                      & = N \left[(x-1) \log\frac{x}{x-1} - 1 \right]
 \end{array}
\]

This is the simplest form to compute the collision
probability as a function of \(N\).

\hypertarget{error-rate.}{%
\subsubsection{Error Rate.}\label{error-rate.}}

The collision probability can never be exactly equal to zero. Instead,
we can guarantee that this probability is less than a limit value
\(\alpha\) which is the tolerated error rate.

\[\alpha > P(\text{Collision})\]
\[1 - \alpha \leq P(\text{No Collision})\]

With a fixed \(N\) and \(\alpha\), we can search for the smallest \(x\)
which satisfies the condition:

\[\log(1-\alpha) < N \left[(x-1) \log\frac{x}{x-1} - 1\right]\]

We can re-write the \(\log\) expression as
\(\log\frac{x}{x-1} = \log \left(1 + \frac{1}{x-1}\right)\), and use the
log approximation when \(\frac{1}{x-1} \rightarrow 0\). The expression
is simplified for large \(x\) as:

\[
\begin{array}{ll}
  \log(1 - \alpha) &< N \left[(x-1)(\frac{1}{x-1} - \frac{1}{2(x-1)^2}) - 1\right] \\
   &< -\frac{N}{2(x-1)}
 \end{array}
\]

Re-arranging the terms to isolate \(x\), we get:

\begin{equation}\label{eq:error_alpha_1}
    x > -\frac{N}{2\log(1-\alpha)} +1
\end{equation}

\hypertarget{analysis}{%
\subsubsection{Analysis}\label{analysis}}

Usually, the FNIR or FPIR of \(5\%\) can be tolerated for non-critical
applications, but is unsuitable for high stake applications such as
financial services. Nevertheless, \(\alpha\) is neither the FNIR nor the
FPIR. FPIR and FNIR are error rates \emph{per transaction}. In other
words, the larger the number of users, the larger the number of errors.
Here, \(\alpha\) is the probability of having at least two
identical templates on the enrolled population.
Therefore, the probability \emph{per user} would be approximately
\(\frac{\alpha}{N} \approx \frac{1 - P(\text{No collision})}{N}\).
Therefore, even if \(\alpha\) is not very low (a few percent), the
collision rate per transaction \(\frac{\alpha}{N}\) would be relatively
small.

If we look at Eq. \ref{eq:error_alpha_1}, the term
\(-\frac{1}{2 \log(1-\alpha)}\) is greater than \(1\) for
\(\alpha < 40\%\). When \(\alpha= 40\%\), it means that in the good case
(\(60\%\) chance), there is no error at all; in the worst case, there
are one or two (or a little bit more) collisions between templates.
For a population of one million, the \emph{individual} error rate is about \(10^{-6}\) which is acceptable.
Accepting \(\alpha = 40\%\), this gives us a simple estimate for the
minimal requirements for \(x\):

\[x_{\alpha=0.4} = - \frac{N}{2 \log(1-0.4)} + 1 \approx N + 1 > N\]

In other words, if \(\alpha < \alpha_{0.4}\), then
\(T > T_{\alpha=0.4} = x_{0.4} N > N^2\). In terms of bits, for a
population of \(N = 2^k\) users, we need at least \(2k\) bits as
\(N^2 = \left(2^k\right)^2 = 2^{2k}\).

\hypertarget{anti-collision}{%
\subsubsection{Anti-Collision}\label{anti-collision}}

How many additional bits are needed to reduce even more the collision
probability ? We can re-write the previous equation Eq.
\ref{eq:error_alpha_1} to get the number of bits as a function of the
tolerance level, using the fact that for small \(\alpha\),
\(\log(1-\alpha) \approx - \alpha\):

\[x > \frac{N}{2 \alpha} + 1\]

Therefore, \(T_{\alpha} > \frac{N^2}{2\alpha}\). This represents in
terms of bits:

\begin{equation}\label{eq:bits_alpha}
    \log_2(T_{\alpha}) = \log_2(2^{k_{\alpha}}) = k_{\alpha} > 2 \log_2(N) - 1 - \log_2(\alpha)
\end{equation}

In this relationship, we see that whatever the population size,
the desired accuracy is independent of the considered population. As a
numerical application with \(\alpha = 10^{-4}\), which is a reasonable
number to prevent any collision from occurring, we need \(13\) extra
bits to provide such a security margin. If we compute the disk space
necessary for a population \(N=10^{10}\) for \(\alpha=0.4\) and
\(\alpha=10^{-4}\), we need:

\begin{itemize}
\item
  \(k_{0.4} = 66\) bits, leading to a total space of \(76.8\) Gigabytes,
\item
  \(k_{10^{-4}} = 79\) bits, leading to a total space of \(92.0\)
  Gigabytes.
\end{itemize}

The difference in disk space between \(\alpha=0.4\) and
\(\alpha=10^{-4}\) is relatively small, as most of the cost is due to
the large number of users.

\hypertarget{conclusion}{%
\subsection{Conclusion}\label{conclusion}}

In this first section, we explored the Birthday Paradox, and computed an
approximation for large numbers. We demonstrated that the number of
possible patterns should be at least \(T = N^2\) to prevent frequent
collisions between templates from occurring, which corresponds to a
minimum of \(2 \log_2(N)\) bits, which has a small impact in terms of
space. We also analyzed  the cost of additional security, as when
\(T = N^2\), a few collisions are still possible. Adding a security
margin requires only a few extra bits, where their number is independent
of the population. This opens the possibilities for large-scale biometry,
as the template size can be small while providing high
distinguishability.

\hypertarget{matching-with-noise}{%
\section{\texorpdfstring{Matching with Noise
\label{sec:noise}}{Matching with Noise }}\label{matching-with-noise}}

In the previous section, we studied the ideal case, where there is no
noise and bits follow a Bernoulli distribution with parameter \(1/2\)
and are independent.

In biometry, these assumptions do not often hold. Especially for noise:
when placing a biometric attribute over a sensor, the outcome is almost
always different, because of factors such as:

\begin{itemize}
\item
  the position is never exactly the same (translation or rotation)
\item
  the quality of the capture may vary with many parameters, like
  temperature, dust, light, humidity, \ldots.
\item
  some internal biological changes impact the attribute (disease,
  physical activity, injury, stress, \ldots)
\end{itemize}

Therefore, it is almost impossible to capture twice the same
sample.

In this section, we propose to study the impact of noise on
recognition accuracy.

\hypertarget{noise-assumption}{%
\subsection{Noise Assumption}\label{noise-assumption}}

\hypertarget{probe-vs-template}{%
\subsubsection{Probe VS Template}\label{probe-vs-template}}

In biometry, we distinguish a probe from a template: a template is a
processed capture stored in a database, while a probe is a
freshly processed capture to compare against stored templates.
When described this way, there is no difference between a template and a probe.

In practice, a single capture is often noisy, and it is better to ask
the user several times to present their biometric attribute to reduce
the noise level. It is possible to ask the user to do
so for enrollment, as enrollment is done only once. However, as we want the
identification to be seamless, only one capture is tolerated for user
convenience. Therefore, we consider that a template is ``free of noise''
(which is not exactly true, as there is always residual noise), while a
probe is always noisy.

\hypertarget{probability-distributions}{%
\subsubsection{Probability
Distributions}\label{probability-distributions}}

We still suppose that bits follow a Bernoulli distribution of parameter
\(1/2\) and are uncorrelated. We assume that probes are affected by
noise with level \(2p\). We put a factor \(2\) to consider bits that
are affected by noise but unchanged. On average, \(p\) will be flipped
(i.e., \(0 \rightarrow 1\) and \(1 \rightarrow 0\)), while a proportion
\(p\) will stay at its initial value (\(0 \rightarrow 0\),
\(1 \rightarrow 1\)). For the following, it is easier to think in terms
of ``flip'' probability \(p\), ranging from \(0\) to \(0.5\), as the
interval \([0.5, 1]\) is symmetric. In the displayed graphics, we
will print \(2p\), to present results over an interval ranging from
\(0\) to \(1\).

The intra-distribution of Hamming distances follows a Binomial
distribution of parameter \((k, p)\), where \(k\) is the length of the
probe (and template).
In contrast, the inter-distribution follows a binomial
distribution of parameter \((k, 1/2)\), as random noise effects
compensate.

\hypertarget{recognition-probability}{%
\subsection{Recognition Probability}\label{recognition-probability}}

To measure the overall probability of making an identification error, we
first need to look at the probability of recognizing one specific user.
If there are \(n\) templates in the database, we need to compute the
Hamming distance between the probe and the \(n\) templates, and select
the identity of the template with the lowest distance. For simplicity,
we suppose there is only one template with the lowest distance,
i.e., no ex-aequo for the winning place.

We denote the distance between the probe and the corresponding template
\(d\). It ranges between \(0\) (no error) and \(k\) (only errors), but
 the correct template would never be recognized in the latter case.
To be recognized correctly, the intra-distance \(d\) can be greater than
\(0\), as long as the distance between the probe and the other
non-matching templates is larger than \(d\).

The probability that the distance between the probe and the
corresponding template is equal to \(d\) is:

\[P(D = d; p) = \binom{k}{d} p^d (1-p)^{k-d}\]

The probability that the distance between the probe and another
non-matching template is greater than \(d\) is:

\[P(D \geq d+1; \frac{1}{2}) = \frac{1}{2^k} \sum_{i=d+1}^k \binom{k}{i+1}\]

Assembling these two previous equations, the probability that the user
(denoted \(u\)) is correctly recognized is:

\begin{equation}\label{eq:noise_proba}
  \begin{array}{ll}
    P(u; p, n)&= \sum_{d=0}^{k-1} \left[ P(D = d; p)  \times P(D \geq d+1; \frac{1}{2})^{n-1} \right] \\
        &= \sum_{d=0}^{k-1} \left[\binom{k}{d}p^d (1-p)^{k-d}  \times \left(\frac{1}{2^k}\sum_{i=d+1}^k \binom{k}{i} \right)^{n-1} \right]
   \end{array}
\end{equation}

The power \(n-1\) considers the other patterns against
which the distance should not be lower or equal to \(d\).

\hypertarget{group-recognition}{%
\subsection{Group Recognition}\label{group-recognition}}

In the previous paragraph, we looked at the probability of
correctly identifying one user among a group. Now, we want to
find the probability that \emph{all} users \(u_1, u_2, ..., u_n\) are
correctly recognized. We can express the global probability using the
Bayes decomposition rule:

\[
\begin{array}{ll}
  P(\text{Accept All}) &= P(u_1, u_2, ..., u_N) \\
        &= P(u_1)P(u_2, ..., u_N| u_1) \\
        &= P(u_1)P(u_2|u_1)P(u_3, ..., u_N | u_1, u_2)
\end{array}
\]

Here, we consider that an identification outcome depends on the previous
ones. If \(u_1, u_2, ..., u_i\) have been correctly identified so far,
then \(u_{i+1}\) cannot be mistaken as one of these previously
identified users, as errors are symmetric. This only holds because we
consider the case where there is no identification error at all.
Therefore, the overall acceptance probability is:

\begin{equation}\label{eq:noise_proba_all}
  \begin{array}{ll}
    P(\text{Accept all}, p, N) &= \prod_{i=1}^N P(\text{One user recognized among } i \text{ users}) \\
        &= \prod_{i=1}^N P(u; p, i) \\
        &= \prod_{i=1}^N  \sum_{d=0}^{k-1} P(D = d; p) \times P(D \geq d+1; \frac{1}{2})^{i-1}
  \end{array}
\end{equation}

\hypertarget{case-where-p-0}{%
\subsubsection{\texorpdfstring{Case where
\(p = 0\)}{Case where p = 0}}\label{case-where-p-0}}

When there is no noise, we should get the same results as
in the previous part. The distance between the probe and the
corresponding template must be exactly \(0\). Therefore, the sum in Eq.
\ref{eq:noise_proba_all} can be simplified as only the first term
\(P(D = d; p)\) is equal to \(1\) while all others are equal to \(0\). Next, the
inter-probability is obtained by using the complementary event
\(P(D \geq 1; \frac{1}{2}) = 1 - P(D = 0; \frac{1}{2})\):

\begin{equation}\label{eq:proba_zero}
\begin{array}{ll}
  P(\text{Accept all}; p=0, N) &= \prod_{i=1}^N 1 \times P(D > 0; p=\frac{1}{2})^{N-i-1} \\
                   &= \left(P(D > 0; \frac{1}{2})\right)^{\frac{N^2-N}{2}} \\
                   &= \left(1 - P(D  = 0; \frac{1}{2})\right)^{\frac{N^2-N}{2}} \\
                   &= \left(1 - \frac{1}{2^k}\right)^{\frac{N^2-N}{2}}
  \end{array}
\end{equation}

Now, we can re-arrange Eq. \ref{eq:proba_zero} to fit the organization
of Eq. \ref{eq:error_alpha_1} to verify if both expressions lead to the
same behavior.

First, we can replace \(2^k\) by \(T\) the number of possible templates
and move to the \(\log\) scale:

\[\log(P(\text{accept})) = \log(1-\alpha)= \frac{N^2-N}{2} \log (1 - \frac{1}{T})\]

Then, the log part is approximated as \(\frac{1}{T}\) is sufficiently
small:

\[\log(P(\text{accept})) \approx -\frac{N^2-N}{2} \times \frac{1}{T}\]

And last, we can express \(T\) as a function of \(N\), such as \(T=xN\)

\begin{equation}\label{eq:proba_cmp}
    \log(P(\text{accept})) \approx -\frac{N^2-N}{2xN} = - \frac{N - 1}{2x}
\end{equation}

This expression is very close to the inequality in Eq.
\ref{eq:error_alpha_1}, where the \(N-1\) replaces the \(N\) (which has
a negligible impact as \(N \gg 1\)), and the denominator is \(2x\) in
Eq. \ref{eq:proba_cmp} instead of \(2(x-1)\), where the difference does
not matter either as \(x\) is large.

\hypertarget{how-to-compute-for-p-0}{%
\subsection{\texorpdfstring{How to Compute for \(p > 0\)
?}{How to Compute for p \textgreater{} 0 ?}}\label{how-to-compute-for-p-0}}

When \(p>0\), the expression cannot be simplified. We can re-write Eq.
\ref{eq:noise_proba_all} as:

\begin{equation}\label{eq:product}
  P(\text{Accept all}) = \prod_{i=1}^N \sum_{d=0}^{k-1} w_d \times a_d^{i-1}
\end{equation}

where:

\begin{itemize}
\item
  \(w_d = \binom{k}{d}p^d(1-p)^{k-d}\) with
  \(\sum_{d=0}^{k-1} w_d = 1\).
\item
  \(a_d = \frac{1}{2^k}\sum_{i=d+1}^k \binom{d}{i}\) with
  \(0\leq a_d \leq 1\).
\end{itemize}

The terms \(w_d\) and \(a_d\) are the same whatever the value of \(i\).
The main problem is the term \(a_d^{i-1}\) where the power makes
simplification difficult.

We can provide an upper and lower bound to this value. First, we define
\(f(i)\) as \(f(i) = \sum_{d=0}^{k-1} w_d a_d^{i-1}\), which simplify
Eq. \ref{eq:product} into \(P(\text{Accept all}) = \prod_{i=1}^N f(i)\).
Next, as \(a_d \leq 1\), if \(i < j\), then \(f(i) \geq f(j)\).
Therefore:

\[f(N)^N \leq P(\text{Accept all}) \leq f(1)^N\]

These bounds are too coarse to be useful. Keeping this approximation
idea in mind, let's say that we cut into \(z\) intervals our \(N\)
users, such as \([n_0=1, n_1, n_2, ..., n_z = N | n_i < n_{i+1}]\). We
can re-write the product into multiple sub-products:

\[P(\text{Accept all}) = \prod_{i=0}^{z-1} \prod_{j=n_i}^{n_{i+1}} f(j)\]

Given the previous trick, we can provide a more accurate bounding:
\begin{equation}\label{eq:approx_product}
	\prod_{i=0}^{d-1} f(n_{i+1})^{n_{i+1}-n_i} \leq P(\text{Accept all}) \leq \prod_{i=0}^{d-1} f(n_i)^{n_{i+1}-n_i}
\end{equation}

\hypertarget{numerical-tests}{%
\subsubsection{Numerical Tests}\label{numerical-tests}}

In the following table Table \ref{tab:approx}, we reported the exact value
\(v_{truth} = P(\text{Accept alll})\) and compared it to the lower bound
\(v_{low}\) and upper bound \(v_{high}\). The approximations were done
using \(100\) intervals of equal size (± 1 element).

\begin{table}[h]
  \centering
  \caption{Approximation error done with Eq. \ref{eq:approx_product}}\label{tab:approx}
  \begin{tabular}{|l|r|c|c|c|c|c|c|}
    \hline
\(k\) & \(n\) & \(p\) & \(v_{low}\) & \(v_{truth}\) & \(v_{high}\) &
\(\Delta = \frac{v_{high}-v_{low}}{2 v_{truth}}\) \\
\hline
20 & \(1,000\) & 0.001 & \(0.559\) & \(0.562\) & \(0.566\) & 0.58
\% \\
25 & \(1,000\) & 0.001 & \(0.98\) & \(0.98\) & \(0.981\) & 0.02
\% \\
20 & \(1,000\) & 0.01 & \(0.161\) & \(0.164\) & \(0.167\) & 1.8
\% \\
25 & \(1,000\) & 0.01 & \(0.904\) & \(0.905\) & \(0.906\) & 0.1
\% \\
30 & \(1,000\) & 0.01 & \(0.994\) & \(0.994\) & \(0.995\) & 0.00556
\% \\
50 & \(1,000\) & 0.2 & \(0.467\) & \(0.47\) & \(0.474\) & 0.709
\% \\
150 & \(10,000\) & 0.4 & \(0.615\) & \(0.618\) & \(0.621\) & 0.451
\% \\
35 & \(100,000\) & 0.01 & \(0.0464\) & \(0.0478\) & \(0.0493\) & 3.06
\% \\
35 & \(100,000\) & 0.001 & \(0.779\) & \(0.781\) & \(0.783\) & 0.25
\% \\
35 & \(100,000\) & 0.0001 & \(0.855\) & \(0.857\) & \(0.858\) & 0.156
\% \\
45 & \(1,000,000\) & 0.01 & \(0.369\) & \(0.373\) & \(0.377\) & 0.994
\%\\
\hline

  \end{tabular}
\end{table}

As you can see, the absolute and relative errors are relatively low.
Therefore, we consider this approximation as accurate enough for our
application.

In the next part, we will investigate the behavior in the open world,
where one part of the population is unenrolled.

\pagebreak

\hypertarget{open-world}{%
\section{Open-World}\label{open-world}}

In the two previous sections, we studied the case where all users are
enrolled. Therefore, there cannot be false positives. In real life,
there are many people unknown from the biometric system. Therefore,
identification should first decide between two main cases: is the person
enrolled or not? If the person is enrolled, an identity is returned.

It is common to set a decision threshold \(thr\)to make this initial decision:
if the distance between the probe and the closest template is
less or equal to \(thr\), then the system considers the user enrolled.

\hypertarget{error-rates}{%
\subsection{Error Rates}\label{error-rates}}

\hypertarget{fnir}{%
\subsubsection{FNIR}\label{fnir}}

Because of the threshold, an enrolled user will sometimes not be
recognized appropriately (\(FNIR\)). There are two complementary
sub-cases:

\begin{enumerate}
\def\labelenumi{\arabic{enumi}.}
\item
  No identity is returned (\(FNIR_n\))
\item
  An identity is returned, but not the correct one (\(FNIR_i\))
\end{enumerate}

When no identity is returned,  the distance between the
probe and any template (the correct one and all \(N-1\) incorrect
others) is greater than \(thr\). It is computed as:

\[FNIR_n = P(D > thr, p) \times P\left(D > thr, \frac{1}{2}\right)^{N-1}\]

The second case is the confusion probability, which is the probability
of taking someone for someone else. This is when \emph{at least}
one incorrect template is closer than the correct one and this distance
is lower than the threshold.

\[FNIR_i = \sum_{d=1}^{thr} P(D=d, p) \times \left(1 - P\left(D > d, \frac{1}{2}\right)^{N-1}\right)\]

\hypertarget{fpir}{%
\subsubsection{FPIR}\label{fpir}}

In the open world setup, there are unenrolled people that the system
will try to identify. We get a false positive match if the distance
between the probe and \emph{at least} one template is under the decision
threshold:

\[FPIR = 1 - P\left(D>thr, \frac{1}{2}\right)^N\]

Compared to the two other error rates, this one is independent of the
noise level.

\hypertarget{trade-off-between-parameteres}{%
\subsection{Trade-Off Between
Parameters}\label{trade-off-between-parameteres}}

We have two input variables \(p\) and \(N\), three constraints, and we
can play on two parameters \(k\) and \(thr\). In these paragraphs, we
will describe the methodology for setting \(k\) and \(thr\) efficiently.

\hypertarget{fnir_n}{%
\subsubsection{\texorpdfstring{\(FNIR_n\)}{FNIR\_n}}\label{fnir_n}}

This error rate depends on \(p\), \(N\), \(k\) and \(thr\). Therefore,
it is difficult to optimize this value due to the dependencies.
The \(FNIR_n\) is made of two parts: one which concerns the distance
between the probe and the correct template, and the second concerns the
relationship between the probe and all other templates. Because both are
probability terms, they are bounded by \(1\).
By setting the second term to \(1\), an upper bound
for this error rate is:

\[FNIR_n < P(D > thr, p)\]

Here, this probability is still dependent on the three different variables (as \(thr\) depends on \(k\)).
Nevertheless, this expression is much simpler to compute.
If we denote \(\beta\) the maximal acceptable \(FNIR_n\), we should search for a triplet \((thr, p, k)\) that satisfy:

\begin{equation}\label{eq:fnir_n}
	FNIR_n < P(D > thr, p) \leq \beta
\end{equation}


The \(FNIR_n\) is a \emph{convenience} rate as nothing is at stake. If
the user is not recognized the first time, they can retry again.
Therefore, this rate can be relatively high compared to the two other
rates, and a maximal error rate of \(5 \%\) can be acceptable in
real-world scenarios.

\hypertarget{fnir_i}{%
\subsubsection{\texorpdfstring{\(\mathbf{FNIR_i}\)}{FNIR\_i}}\label{fnir_i}}

The \(FNIR_i\) is the confusion rate, i.e., the probability of taking
someone for someone else. This rate depends on the four variables,
making it difficult to optimize. As for the \(FNIR_n\), we can provide
an upper bound to simplify the parameter selection process. As the
larger the threshold is, the greater the \(FNIR_i\), we set \(thr=k\) to
remove the threshold dependency:

\[FNIR_i = FNIR_i(thr) < FNIR_i(thr=k) = FNIR_i^{\infty}\]

The \(FNIR_i\) is the confusion rate \emph{per transaction}. Therefore,
the number of errors scales linearly with the number of users trying to
identify. However, we may want to guarantee a fixed error rate
independent of the number of users. This was the topic of the two
previous sections \ref{sec:nonoise} and \ref{sec:noise} where the
global error rate can be guaranteed.

In section \ref{sec:nonoise}, the minimal requirement for no noise is \(k > 2 \log_2(N)\). For a larger noise level, the
necessary \(k\) needs to be computed case by case.

\hypertarget{fpir-1}{%
\subsubsection{\texorpdfstring{\(\mathbf{FPIR}\)}{FPIR}}\label{fpir_i}}

The last error rate to look at is the \(FPIR\), which is independent of
the noise level \(p\). The two \(FNIR\) rates provide us the lowest acceptable
values for \(k\) and \(thr\). When \(thr\) increases, the \(FPIR\)
increases too, while when \(k\) increases, the \(FPIR\) decreases.
Therefore, given the minimal \(k\) provided by the \(FNIR_i\), we need
to increase it until the \(FPIR\) conditions are met.

\[
\begin{array}{ll}
  FPIR &= 1 - (1 - P(D \leq thr, \frac{1}{2}))^N \\
        &\approx 1 - (1 - N \times P(D \leq thr, \frac{1}{2})) \\
        &\approx N \times P(D \leq thr, \frac{1}{2})
  \end{array}
\]

The \(FPIR\) is an error rate \emph{per transaction}. The larger the
unenrolled population \(N^-\), the greater the number of errors.

We need to pay attention to the deployment scenario. For attended
biometry, i.e., where the biometry cannot be captured without the user's
consent, the number of unenrolled persons that will try to authenticate
themselves is relatively low. However, for unattended biometry, where
the system will try to recognize any person (like video cameras), the
number of unenrolled persons will be very large, maybe larger than the
enrolled population.

The target system probability could be:

\[
\begin{array}{ll}
  P\left(\text{One error over } N^- \text{transactions}\right) &= 1 - (1 - FPIR)^{N^-} \\
  &\approx 1 - (1 - N \times P(D \leq thr, \frac{1}{2}))^{N^-} \\
  &\approx 1 - (1 - N N^- \times P(D \leq thr, \frac{1}{2})) \\
  &\approx  N N^- \times P(D \leq thr, \frac{1}{2})
  \end{array}
\]

As for the \(FNIR_n\), we can approximate \(P(D \leq thr, \frac{1}{2})\)
using a normal law \(\mathcal{N}(k/2, \sqrt{\frac{k}{4}})\). Setting
\(\gamma\) as the maximal error rate for
\(P\left(\text{One error over } N^- \text{transactions}\right)\), we
have:

\begin{equation}\label{eq:FPIR_approx}
    thr < \frac{k}{2} - f^{-1}\left(\frac{\gamma}{N N^-}\right) \sqrt{\frac{k}{4}}
\end{equation}

where \(f^{-1}(.)\) satisfies \(\gamma = P\left(Y > f^{-1}(\gamma) \right) | Y \sim \mathcal{N}(0, 1)\).

\hypertarget{search-procedure}{%
\subsection{Search Procedure}\label{search-procedure}}

We can summarize the search procedure for \(k\) and \(thr\) by Algo.
\ref{alg:search}.

\begin{algorithm}
\caption{Search Procedure}\label{alg:search}
    \begin{algorithmic}
    \Require $\alpha, \beta, \gamma, N, N^-, p$
    \State $k \gets g_1(\alpha, N, p)$ \Comment{Using Eq. \ref{eq:product}}
    \State $flag \gets \text{true}$
    \While{$flag$}
        \State $thr_0 \gets g_2(\beta, N, p, k)$ \Comment{Using Eq. \ref{eq:fnir_n}}
        \State $thr_1 \gets g_3(\gamma, N, N^-, k)$ \Comment{Using Eq. \ref{eq:FPIR_approx}}

        \If{$thr_0 < thr_1$}
            \State $flag \gets \text{false}$
        \Else
            \State $k  \gets k+1$
        \EndIf
    \EndWhile

\Return $(k, thr_0)$
    \end{algorithmic}
\end{algorithm}

\hypertarget{conclusion-1}{%
\subsection{Conclusion}\label{conclusion-1}}

In this section, we have provided bounds and approximations for the
three error rates in the case of an open-world situation. To satisfy
these three conditions, a simple search procedure can be followed.

\pagebreak

\hypertarget{numerical-results}{%
\section{Numerical Results}\label{numerical-results}}

In this part, we present numerical evaluations by displaying
corresponding curves. In the first subpart, we study the evolution of
\(P(\text{Accept})\) as a function of the other parameters \(k\), \(p\)
and \(N\) the population size. In the second subpart, we fix the
tolerance level \(\alpha\) and study the relationships between
parameters to preserve this error level. This enables us to look at two
parameters at a time. After that, we model the relationships between
noise, population and number of bits. Given these results, we estimate
the database size requirements. Last, we evaluate these requirements in
the open-world case.

\hypertarget{ptextaccept}{%
\subsection{\texorpdfstring{\(P(\text{Accept})\)}{P(\textbackslash text\{Accept\})}}\label{ptextaccept}}

\hypertarget{paccept-fn-pk}{%
\subsubsection{\texorpdfstring{\(P(Accept) = f(N, p;k)\)}{P(Accept) = f(N, p;k)}}\label{paccept-fn-pk}}

In this experiment, we study the overall acceptance probability as a
function of the population size for a fixed number of bits (\(k=40\)).
We study the behavior for doubling the noise level from \(0.15 \%\) to
\(40\%\).

\begin{figure}[H]
\centering
\includegraphics[width=0.7\textwidth]{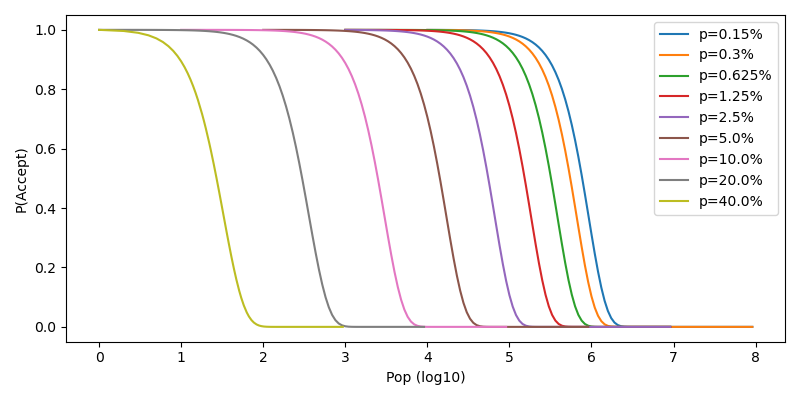}
\caption{Encoding capacity for k=40}
\end{figure}

As expected, the larger the noise level, the smaller the population
we can enroll without having high error rates. At some point, we reach a
ceiling, as for \(k=40\), we can enroll at best one million users
without noise, as \(\log_{10}(2^{40/2}) \approx 6\).

\hypertarget{paccept-fn-k-p}{%
\subsubsection{\texorpdfstring{\(P(Accept) = f(N, k; p)\)}{P(Accept) = f(N, k; p)}}\label{paccept-fn-k-p}}

Compared to the previous experiment, the noise level is fixed to
\(10 \%\) while the number of bits is changed from \(10\) to \(50\)
bits.

\begin{figure}[H]
\centering
\includegraphics[width=0.7\textwidth]{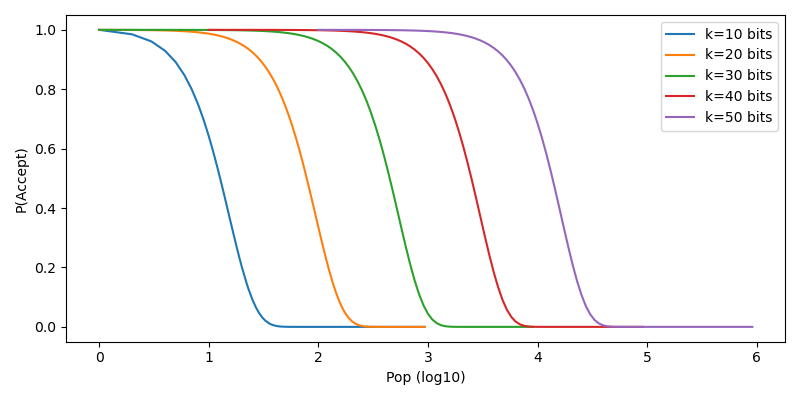}
\caption{Encoding capacity for \(p=10 \%\)}
\end{figure}

Even under noisy conditions, it seems that \(k\) additional bits increase
the capacity from \(N\) to \(N \times K\), where \(K\) is independent of
\(N\).
This result was proved for the no-noise case, but we could not demonstrate this  due to the complexity of the noisy case.
The next subsections will empirically study this relationship between bits and population.

\hypertarget{paccept-fk-p-n}{%
\subsubsection{\texorpdfstring{\(P(Accept) = f(k, p; N)\)}{P(Accept) = f(k, p; N)}}\label{paccept-fk-p-n}}

Last, we study the resilience behavior when adding bits while increasing
noise for a fixed-size population, where \(N = 10^6\). Here, the
noise is increased by adding \(5\%\) for each step, ranging from \(5\%\)
to \(40\%\).

\begin{figure}[H]
\centering
\includegraphics[width=0.7\textwidth]{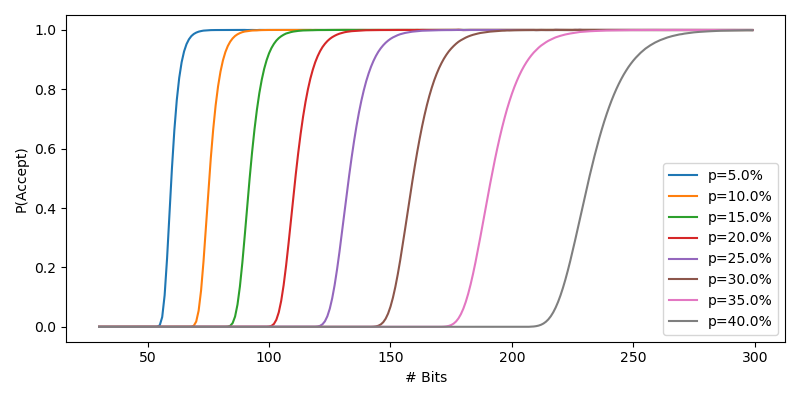}
\caption{Numbers of bits necessary for \(N=10^6\) for different noise levels}
\end{figure}

The relationship between the noise level and the number of bits is not
linear, as we need more bits to compensate for a change from \(35\) to
\(40\%\) of noise compared to a change from \(5\) to \(10\%\).

\hypertarget{k-fn-alpha}{%
\subsection{\texorpdfstring{\(k = f(N; \alpha)\)}{k = f(N; \textbackslash alpha)}}\label{k-fn-alpha}}

In many real-world applications, there is often a target accuracy to
reach, or a maximal error rate. In this part, we fix simultaneously the tolerance level
\(\alpha = 1 - P(\text{Accept})\) to study two parameters.

In this experiment, we study the relationship between the number of bits
and the population size and display several curves for different noise
levels. We repeated the experiment for two tolerance levels, one with
\(\alpha=0.5\) and the other with \(\alpha=0.1\).

\begin{figure}[H]
\centering
\includegraphics[width=0.7\textwidth]{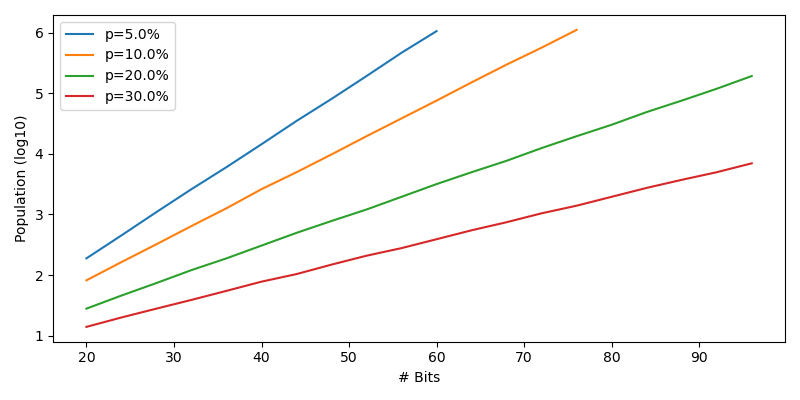}
\caption{Tolerance level \((1-\alpha) > 0.5\)}
\end{figure}

\begin{figure}[H]
\centering
\includegraphics[width=0.7\textwidth]{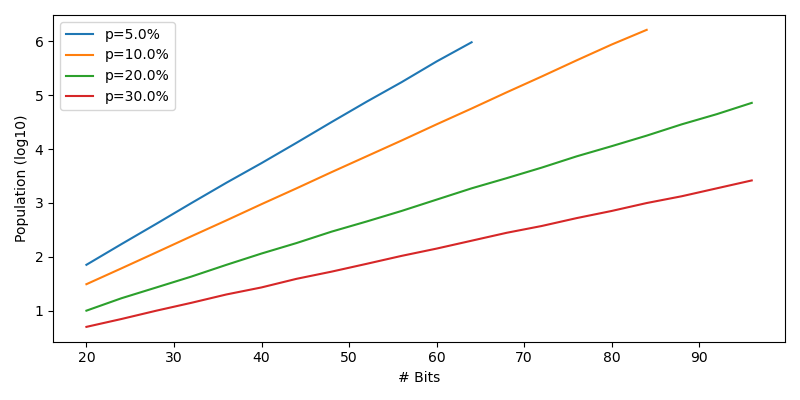}
\caption{Tolerance level \((1-\alpha) > 0.9\)}
\end{figure}

We have a clear linear relationship between the population log and the
number of bits, whatever the tolerance level \(\alpha\) and the noise
level \(p\). When fitting a linear model with the form \(ax + b\), the
fit is very good with a \(r^2 > 0.99\). The only differences between the
two figures are the coefficients \(a\) and \(b\), which depend on
\(\alpha\) (and \(p\)).

\hypertarget{getting-slope-coefficients-as-a-function-of-noise}{%
\subsection{Getting Slope Coefficients as a Function of
Noise}\label{getting-slope-coefficients-as-a-function-of-noise}}

From the previous plots, we can assume that the number of bits is
linearly linked to the log population size \(N\). Therefore, we can
write that:

\[\log_2(N) = a_p \times k + b_p\]

In practice, we are more interested in the reciprocal relationship:
Given a population of \(N\), we want to know how many bits are necessary
to prevent collision. In that case, we write the relationship as:

\[k = A_p \times \log_2(N) + B_p\]

where \(A_p = \frac{1}{a_p}\) and \(B_p = -\frac{b_p}{a_p}\).

We collected different coefficients \(A_p\) and \(B_p\) by increasing the
noise level, increasing the noise by \(1\%\) from \(1\%\) to \(50\%\),
and fixed \(\alpha\) to \(10^{-4}\).

We limited the exploration to noise levels between \([0 \%, 50\%]\) for
several reasons. We expect the noise in real life to be smaller than
\(50\%\); otherwise the capture is very inaccurate. In
that case, it might be better to optimize sensor quality rather than
invest in larger databases and a slower matching system.
Next, even if a larger level of noise were studied,
splitting the data into different intervals and fitting them separately
would have been preferable, as errors on large values will mask errors on small values.

We obtained the following curves:

\begin{figure}
\centering
\includegraphics[width=\textwidth]{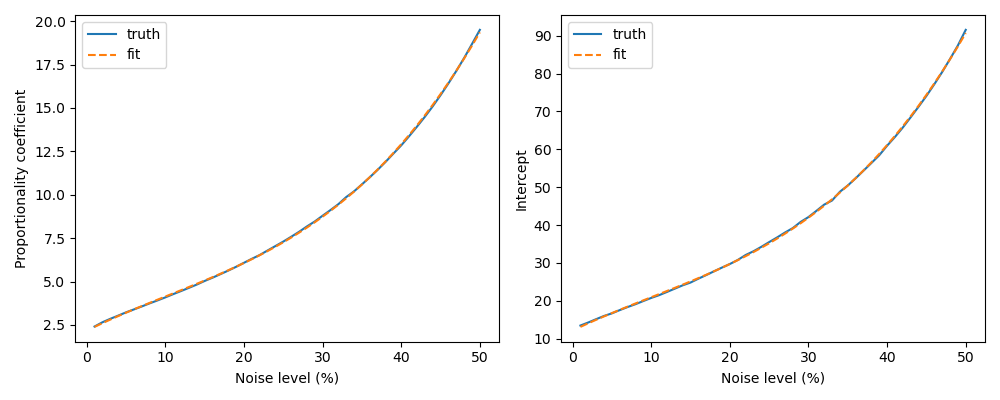}
\caption{Coefficients curves and fit: Left: \(A_p\), right: \(B_p\)}
\end{figure}

As you can see, the larger the noise, the larger the slope and the
intercept in absolute value. We fitted a polynom of degree \(3\) where
the line (orange in the figure) perfectly overlaps the experimental
values, where the coefficients are:

\[A(p) =  1.28 \times 10^2 p^3 - 40.2 p^2 + 22.4 p + 2.17\]
\[B(p) = 5.80 \times 10^2 p^3 - 1.75 p^2 + 99.8 p + 12.1\]

We obtained an MSE of \(0.048\) for coefficients \(A_p\) and \(0.25\) for
coefficients \(B_p\). With this fit, we get \(A(0) = 2.17\), which is
slightly greater than \(2\) and an intercept of \(B(0) = 12.1\) which is
very close to \(13 - 1 = 12\) from Eq. \ref{eq:bits_alpha} when
\(\alpha=10^{-4}\). These differences are acceptable for our application
where we are searching for the right order of magnitude, not necessarily
the exact value with no errors.

Here, the coefficients are expressed as a function of \(\log_2(N)\),
compared to \(\log_{10}(N)\) in the previous choices. In the previous
experiments, we studied the impact of noise and \(k\) on the maximal
capacity. It is much easier in that case to use the \(\log_{10}(N)\),
while it is much more difficult to know by heart  what \(2^x\)
corresponds to. Here, we want a relationship between \(k\) and \(N\). In
the first section, we found that \(T = 2^{2k}\) when \(N = 2^k\).
Therefore, to compare to this first result, we expressed the coefficient
\(A_p\) in base \(2\).

\hypertarget{biometric-database-size}{%
\subsection{Biometric Database Size}\label{biometric-database-size}}

Using the coefficients \(A(p)\) and \(B(p)\), we can estimate the size
required to store templates. Here, we consider a population of ten billion,
where \(10^{10} \approx 2^{33.2}\). The results are listed in the following table.

\begin{table}[h]
  \centering
  \caption{Relationship between noise, bits, and database size.}\label{tab:noise_db}
  \begin{tabular}{|l|r|c|c|c|}
    \hline

Noise level \(p\) (\%) & \(A(p)\) & \(B(p)\) & \(k(p, N)\) & DB size
(GB)\\
    \hline
0 & 2.17 & 12.1 & 85 & 98.95\\
5 & 3.21 & 16.7 & 124 & 144.4\\
10 & 4.14 & 20.9 & 159 & 185.1\\
15 & 5.06 & 25.1 & 194 & 225.8\\
20 & 6.07 & 29.7 & 232 & 270.1\\
25 & 7.26 & 35.2 & 277 & 322.5\\
30 & 8.74 & 41.9 & 333 & 387.7\\
35 & 10.6 & 50.4 & 403 & 469.2\\
40 & 12.9 & 61.1 & 491 & 571.6\\
45 & 15.8 & 74.4 & 600 & 698.5\\
50 & 19.4 & 90.7 & 734 & 854.5\\
\hline

  \end{tabular}
\end{table}

As you can see, even with much noise, the whole Earth's
population can fit in a regular disk. These are encouraging results for
developing identification systems, as storage size is not the
limiting problem.

\hypertarget{open-world-1}{%
\subsection{Open-World}\label{open-world-1}}

So far, we have studied the confusion rate in a closed world where all
the population is enrolled. Now, we can study the impact of a threshold,
and see how it impacts the memory requirements.

For this experiment, we set \(FNIR_n^{\max} = 10^{-2}\),
\(FPIR = 10^{-4}\) and \(P(\text{reject all external}) = 10^{-4}\), as
it seems reasonable values for most scenario. Here, we set
\(N = N^- = 10^{10}\). The rationale of this last choice is the
following: the system should be able to enroll the whole population size
without any conflict (\(N = 10^{10}\)). However, most of the
population would not be enrolled on day one. Therefore, at max, we would have
\(10^{10}\) non-enrolled people (\(N^- = 10^{10}\)).

\begin{table}[h]
  \centering
  \caption{Threshold values for various noise level.}\label{tab:threshold}
  \begin{tabular}{|c|c|c|l|}
    \hline
Noise (\%) & \(k_0\) & \(k\) & \(thr\)\\
    \hline
5 & 125 & 125 & 9\\
10 & 160 & 160 & 16\\
15 & 195 & 195 & 25\\
20 & 233 & 233 & 35\\
25 & 278 & 278 & 49\\
30 & 334 & 334 & 67\\
35 & 404 & 404 & 90\\
40 & 492 & 492 & 120\\
45 & 601 & 601 & 160\\
50 & 735 & 735 & 212\\
\hline
  \end{tabular}
\end{table}

In Table \ref{tab:threshold}, \(k_0\) is the number of bits obtained in the closed case
where there is no threshold, thanks to the coefficient \(A_p\) and
\(B_p\). \(k\) is the adjusted value in case there are the
\(FNIR_n\) and \(FPIR\) constraints.

The results have been obtained by approximating the main parameters
using the normal law approximations. However, due to numerical errors
for low \(p\), we refined the search by a local search to obtain the
exact minimal parameters.

Here, surprisingly, \(k_0 = k\), i.e., no difference exists between
the open-world and closed world memory requirements. This is due to the
high distinguishability requirements, which makes authentication of
outsiders very difficult. When the constraints for the \(FPIR\) are
increased, more bits are necessary for the low noise level, but
there is no change when \(p\) is large.

In conclusion,  using a threshold to prevent unenrolled users from
authenticating has a very limited impact on the memory requirements, as
no extra bits are needed.

\hypertarget{conclusion-2}{%
\section{Conclusion}\label{conclusion-2}}

In this work, we studied the number of bits necessary to store biometric
templates while avoiding collision. As data collection is subject to
noise, we studied its impact in the identification case.

In the ideal case of no noise, the number of bits grows in
\(2 \log_2(N)\). However, more bits are needed
to compensate for uncertainty in the presence of noise. When the noise level is fixed, there is
still a linear relationship between the log of the population and the
number of bits necessary, which eases the estimation of the required
database size. While the noise increases the need for more bits, most
of the cost is associated with the number of enrolled users. When moving
from \(5\%\) to \(50\%\) of noise, we only need to multiply by \(6\) the
disk space needed.

We considered the open-world setup, where one part of the population is
not enrolled yet. We provided the probabilistic formulation of the
\(FPIR\) and \(FNIR_i/FNIR_n\) as a function of the threshold.
Additional bits are needed to adapt to the
threshold case for a small noise level.
However, the threshold has no impact for a large noise level, as the required number of bits is unchanged.

This is an encouraging step towards large-scale identification systems.
Nevertheless, there are still many challenges to solve. First, all
through this work, we assumed that bits are independent. Research is
needed to provide efficient compression algorithms to remove redundancy
from templates. Next is the matching question: how much time is
required, and how can we speed up the identification process as it
requires comparing all the \(n\) stored templates to the probe. Last, we
have the question of cancelable biometrics and secure biometric
templates. As for today, while cancelable biometrics such as Biohashing
~{[}12{]} enables the creation of several protected templates from a single
unprotected one, the matching accuracy is often decreased. We need to
study this system more in detail to provide additional security
guarantees without accuracy loss in this setup.

\clearpage

\pagebreak

\bibliographystyle{unsrtnat}

\bibliography{references}  

\end{document}